# Fine-tuned atomistic foundation model for accurate sodium-carbon chemistry in Na-ion batteries

Andres F. Usuga[2], Ihor Radchenko[2], Yuxin Tang[1,3*], and Oleksandr I. Malyi[1, 2*]

[1]Qingyuan Innovation Laboratory, 1 Xueyuan Road, Quanzhou 362801, P. R. China

[2]Centre of Excellence ENSEMBLE3 Sp. z o. o., Wolczynska Str. 133, 01-919, Warsaw, Poland

[3]College of Chemical Engineering, Fuzhou University, No. 2 Xueyuan Road, Fuzhou, 350116, China

[*]**Corresponding authors:** yxtang@fzu.edu.cn(Y.T) oleksandrmalyi@gmail.com (O.I.M)

## ABSTRACT

Atomistic foundation models have emerged as a promising route to DFT-level accuracy for materials simulation with complex chemical environments. However, their reliability across competing bonding regimes and, more importantly, how complex bonding affects predictions of material properties remain unclear. Among materials with competing bonding regimes, sodium storage in hard carbon poses a substantial challenge. Hard carbon features diverse local structural motifs, including defective and curved graphitic regions, amorphous regions, and closed/open pores. Motivated by its importance for Na-ion batteries, we systematically evaluate existing pretrained foundation models based on the MACE architecture across different local Na/C motifs. We find that the pretrained models fail to accurately capture Na-Na and Na-C interactions for metallic Na clusters on graphene. To recover a DFT-level energetic description, we employ targeted fine-tuning, showing that it can describe the energetic stability of local motifs and achieve sufficient transferability to unseen Na/C environments. With the developed fine-tuned model, we establish a correlation between closed-pore radius in hard carbon and plateau voltage, showing that the small pores contribute at larger voltages than large ones, suggesting thus how the developed fine-tuned foundation model can link atomistic information with materials design for complex bonding environments.

## INTRODUCTION

Na storage in C-based anodes is governed by a competition between chemically distinct interactions that coexist within a single electrode. This is particularly true for hard carbon (one of the most promising anode materials for Na-ion batteries[1,2]), where local structural motifs range from defective and curved graphene and short-range-ordered graphite-like domains to closed/open pores.[3] Each such structural motif contributes differently to the voltage profile:[4,5] adsorption at undercoordinated sites (i.e., defects) usually contributes to the higher-voltage slope region, while intercalation between turbostratic layers[6] and pore-filling are often associated with a plateau-like region (<0.1 V) in a hard carbon electrode (Fig. 1a). This picture implies that to theoretically predict thermodynamics, voltage trends, and/or transport properties across different degrees of charge, one must reproduce not only separate constituents (i.e., bulk Na and the C matrix) but also the continuous transition between different types of involved local interactions. In this light, density functional theory (DFT) represents a state-of-the-art reference,[7–9] at least if non-naïve DFT approximations[10,11] are involved. However, for a hard carbon electrode, DFT performance is often limited by computational cost: due to system complexity and the need to accurately describe the distribution of local motifs, a large supercell, often unrealistically large for DFT, is usually required. Machine-learning interatomic potentials (MLIPs) promise to bridge this gap by delivering near-first-principles accuracy at orders-of-magnitude lower cost. The strength of MLIPs lies not only in describing local chemical environments per atom (lower cost) but also in capturing and learning from diverse configurational chemical spaces, unlike classic force field potentials, which are parametrized to restricted local motifs.[12–15] The accuracy of MLIPs depends on two factors: how well the configurational space is sampled (i.e., whether the dataset contains a high diversity of structures representing the system) and the flexibility of the architecture of the potential, especially for a complex span of local geometries, as possessed by C-based anodes. Although different specific MLIP models have been trained to reproduce specific chemistries of carbon, such as amorphous carbon,[16–18] or sodium storage in hard C-like structures,[19–22] for hard carbon systems, beyond high predictive performance on local environments (i.e., adsorption, intercalation, or pore filling), the model must be able to capture the transition regions. Foundation models have been gaining interest because they may help address this challenge by using datasets with the largest possible quantity of diverse chemical compositions.[23] Specifically, foundation models based on the MACE architecture have recently demonstrated impressive performance across diverse chemistries, raising expectations that they can serve as scalable surrogates for DFT in battery materials modeling and complex chemistries.[23–25] One example demonstrating the significant advance in battery modeling employing MACE-based foundation models is the melt-quench simulation to predict amorphous carbon structures[23] and the accurate simulation of Na-ion intercalation energetics (Figs. 1b, S1, and S2). Although foundation MACE models have demonstrated the capability to capture the amorphous-like region for hard carbon,[23,24] it must also be ensured that they are able to capture the span of local structural motifs on Na/C, having the ability to describe both the graphitic-like and amorphous-like regions and, more importantly, capture the competing bonding interactions in Na/C, where this bonding cannot be described as a simple phenomenon of adsorption; in this case the weakness of this binding interaction is governed by the weak charge transfer from sodium to carbon, and the metallic behavior originating from Na-Na cohesion. Motivated by this, herein, using the representative motifs summarized in Fig. 1a, we demonstrate that some of the existing pretrained MACE models can reproduce parts of the Na-carbon landscape (i.e., Na intercalation chemistry in graphite-like domains, Figs. 1b, S1, and S2), but all of the tested models still misrepresent the metallic-ionic interplay (i.e., formation of a metallic cluster, Fig. 1c) phenomenon, which controls the thermodynamic stability of Na arrangements across the different local motifs presented in hard carbon. These errors increase when the pretrained model attempts to capture the competing bond between Na-Na and Na-C, which determines whether the sodium will stay as dispersed atoms or can aggregate to form metallic Na clusters. These findings highlight a critical point: for complex binding interactions and diverse bonding regimes (metallic-ionic) in systems such as carbon-based electrodes, an accurate foundation model must capture the underlying physics, not just achieve good global accuracy in energy and forces. Motivated by these results, we fine-tune a sodium-carbon foundation model to capture bonding competition accurately and demonstrate its practical application to Na storage in hard-carbon motifs. The computed results show that the targeted fine-tuned potential can not only capture the main energetics of Na-C interactions but also be used to understand the pore-filling process in representative pore-local motifs. For instance, the computed results demonstrate that, while it is proposed that Na pore filling yields a unique plateau voltage[9], we find that different pores contribute to Na storage at different voltages. Increasing pore size increases the number of Na atoms that can be stored inside the pore, but it also reduces the voltage at which the pores are thermodynamically activated for the Na-storage mechanism. In practical terms, this means small pores have a much stronger thermodynamic pore-filling driving force than larger ones.

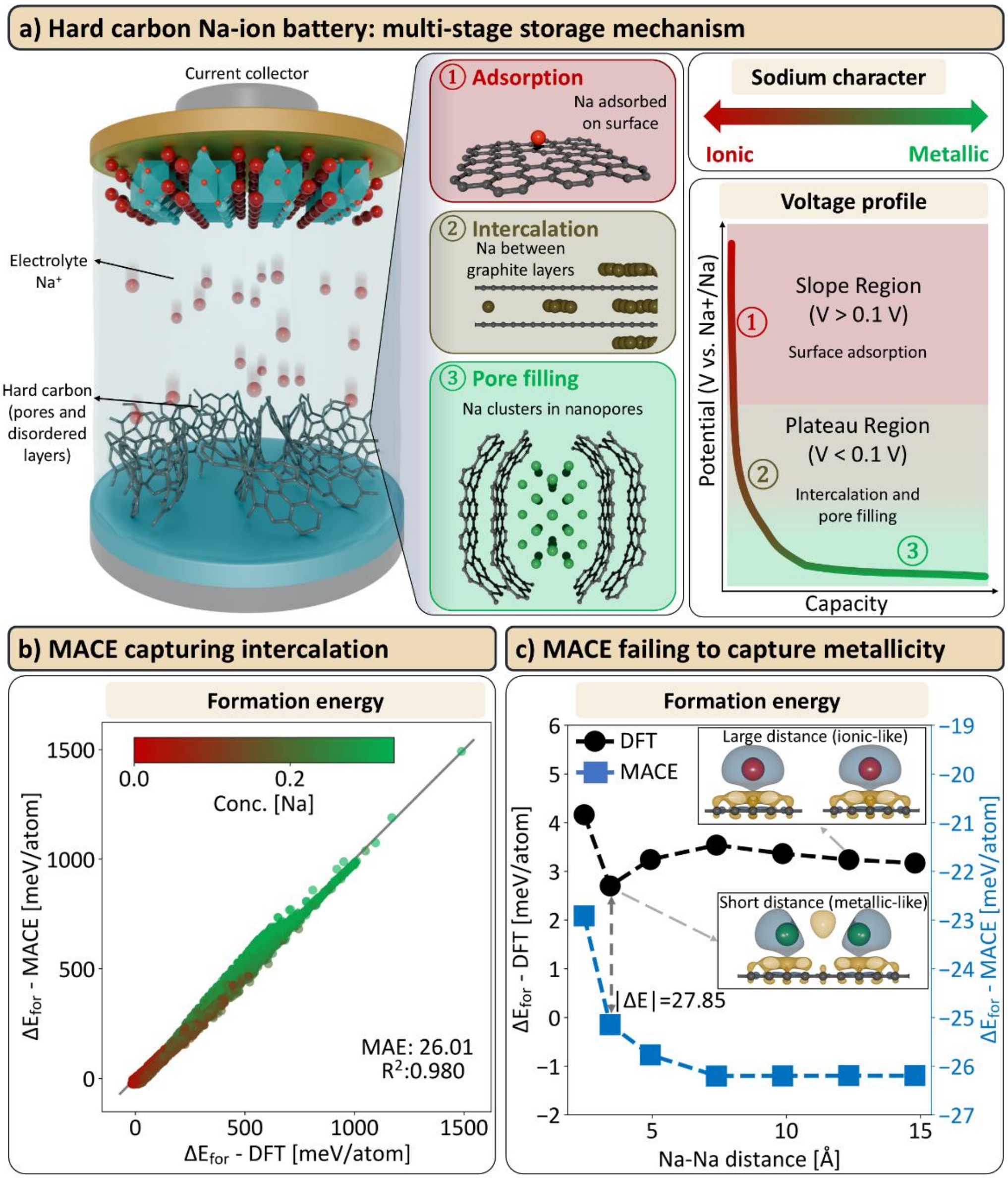

**Figure 1. Na storage mechanism in hard carbon and MACE MATPES ability/inability to capture the main storage mechanism.** (a) Schematic representation of the hard-carbon Na-ion battery, detailing Na storage mechanisms and the voltage profile showing slope (V > 0.1 V) and plateau (V < 0.1 V) regions, along with the expected ionic/metallic character for the corresponding storage mechanism. (b) Parity plot of the formation energy of Na on graphite for the intercalation storage mechanism, comparing the pretrained foundation MATPES model with DFT data; the color scale indicates the atomic Na concentration in each structure; for this dataset, van der Waals dispersion is not included. (c) Formation energy of $Na_2$ in pristine graphene vs. the Na-Na distance; each Na is adsorbed at the center of a 6-ring carbon site; energetic values are without dispersion. Charge density difference plot for two selected distances: a short distance where Na exhibits more metallic character (Na shown in green), and a long distance where ionic character predominates (Na shown in red). The isosurface is obtained from the charge density difference at +0.0002 and -0.0002 e/Å$^3$, corresponding to the acceptance (yellow) and depletion (blue) zones, respectively. In (a) and (c), the stick-and-ball and stick models, drawn in black, are used to represent carbon structures. Polyhedral representations are used to schematically depict a cathode.

## RESULTS AND DISCUSSION

Due to a difference in electronegativities, Na interaction with C is expected to have an ionic character when Na donates electrons to the surrounding C atoms. While this feature is common for battery systems, Na-C bonding is relatively weak: Na is not readily stored on a pristine graphene surface[26] and does not intercalate between graphite layers unless they are expanded.[6] For example, Na intercalation into graphite with fixed interlayer spacing is thermodynamically unfavorable, with positive formation energies of 12.55 and 10.63 meV atom$^{-1}$ for $NaC_6$ and $NaC_{28}$, respectively.[6] This weak Na-C bonding implies that Na storage in carbon cannot be described as a simple adsorption or intercalation process and instead is governed by a competition between two tendencies: (i) charge transfer to the carbon framework and (ii) Na-Na cohesion leading to locally metallic clustering. This interplay is particularly important for the plateau region of the hard carbon voltage profile, where the pore filling via metallic Na cluster formation is believed to be the dominant Na storage mechanism.[10,27–29] While this interplay is accurately captured by DFT, the pretrained foundation model (i.e., MATPES-PBE-0 (MATPES)[30]) misses this stabilization. This is well illustrated by the interaction energy between two Na atoms in Fig. 1c, where DFT shows a local energy minimum, whereas MACE predicts a repulsive interaction. The inability of MACE to track DFT data across different types of Na arrangements (i.e., clustered and dispersed) is even more evident in Fig. 2a,b, which show significant deviations between DFT and pretrained MATPES energetics for various Na clusters on a graphene surface. This misrepresentation is not only attributed to the pretrained MATPES model, but it is also observed for the oldest pretrained models[23] (i.e., MP-0a, MP-0b, and MPA-0; see Fig. S3), including most of the heads of the MH-0 and MH-1 models (see Fig. S4). The pretrained MATPES MACE model shows poor agreement with DFT, with a mean absolute error (MAE) of 32.43 meV/atom and an $R^2$ of -36.32 for the dataset of Na clusters on graphene. This limitation, however, is not expected to reflect a limitation of MACE itself, but rather insufficient representation of these Na/C environments in the database used to train the specific MACE potential. Specifically, the MATPES MACE potential training set is limited to Materials Project[31,32] ground-state calculations, with structures sampled from 300 K molecular dynamics, which provide limited sampling of ionic-metallic interplay within Na-C systems.[30] Moreover, because most structures in the training set are close to the ground state (including those from molecular dynamics simulations), MATPES's ability to describe the complex Na storage mechanism in hard carbon may be limited. Indeed, even for pure graphite systems, the pretrained MATPES MACE model is unable to accurately reproduce the change in the total energy for graphite with AA and AB stacking with respect to the interlayer distance (Fig. 2d). Motivated by these observations and as a *'proof-of-concept'*, we constructed a targeted fine-tuning database designed to capture both Na clustering motifs on graphene (structures from Fig. 2b) and expanded structures of graphite and sodium (including the graphite and sodium reference-state energetics used in the formation-energy analysis). Specifically, (i) we constructed a DFT subset containing Na arrangements on a 10 × 10 × 1 graphene supercell, ensuring that both clustered and dispersed sodium configurations are included (Na-Na bond distances between 3.50 and 7.39 Å); this subset contains 1.985 configurations, and (ii) we included another subset, composed of expanded structures of metallic sodium BCC and graphite with AA and AB stacking, with different factors of expansion, ranging between 0.5 and 1.5 times the equilibrium lattice; this subset contains 300 configurations (for more detailed information, see dataset generation in the SI). We trained the *'proof-of-concept'* fine-tuned model on this targeted dataset. After fine-tuning MATPES-MACE on this combined database (using the same DFT setup as the original pretrained MATPES-MACE), the model reproduces DFT energetics much more accurately. As shown in Fig. 2b, this improvement is substantial. For instance, the MAE decreases from 32.43 to 1.65 meV atom$^{-1}$, while $R^2$ increases from -36.32 to 0.87. Additionally, fine-tuned MATPES models using different seeds (but the same dataset) showed low variability between the predicted formation energies and atomic forces per component (see Figs. S5-S8). More importantly, the fine-tuned model recovers the correct energetic ordering across the areal cluster-size descriptor introduced in Fig. 2a. The Na arrangements correspond to compact Na clusters and a smaller occupied area per Na atom, as their energetic stability is reproduced by the fine-tuned model, in consistent agreement with DFT, while more dispersed Na atoms on graphene remain higher in energy. This indicates that the *'proof-of-concept'* fine-tuned foundation model learned the energetic fingerprint of Na-C and Na-Na bonding, rather than simply reducing numerical error. This is further illustrated in the two-Na-atom limit in Fig. 2c, where the fine-tuned model reproduces the local stabilization associated with short-range Na-Na interaction on graphene. Finally, including graphite AA and AB stacking and sodium structures improves the reference-state energetics in Fig. 2d, reducing the large graphite-spacing-dependent errors of the pretrained model to only a few meV atom$^{-1}$. In graphite, this is especially important because the misrepresentation of the reference state led to cumulative errors in property calculations, affecting thermodynamics and thereby the voltage, and this error can obscure the smaller energetic changes between Na arrangements.

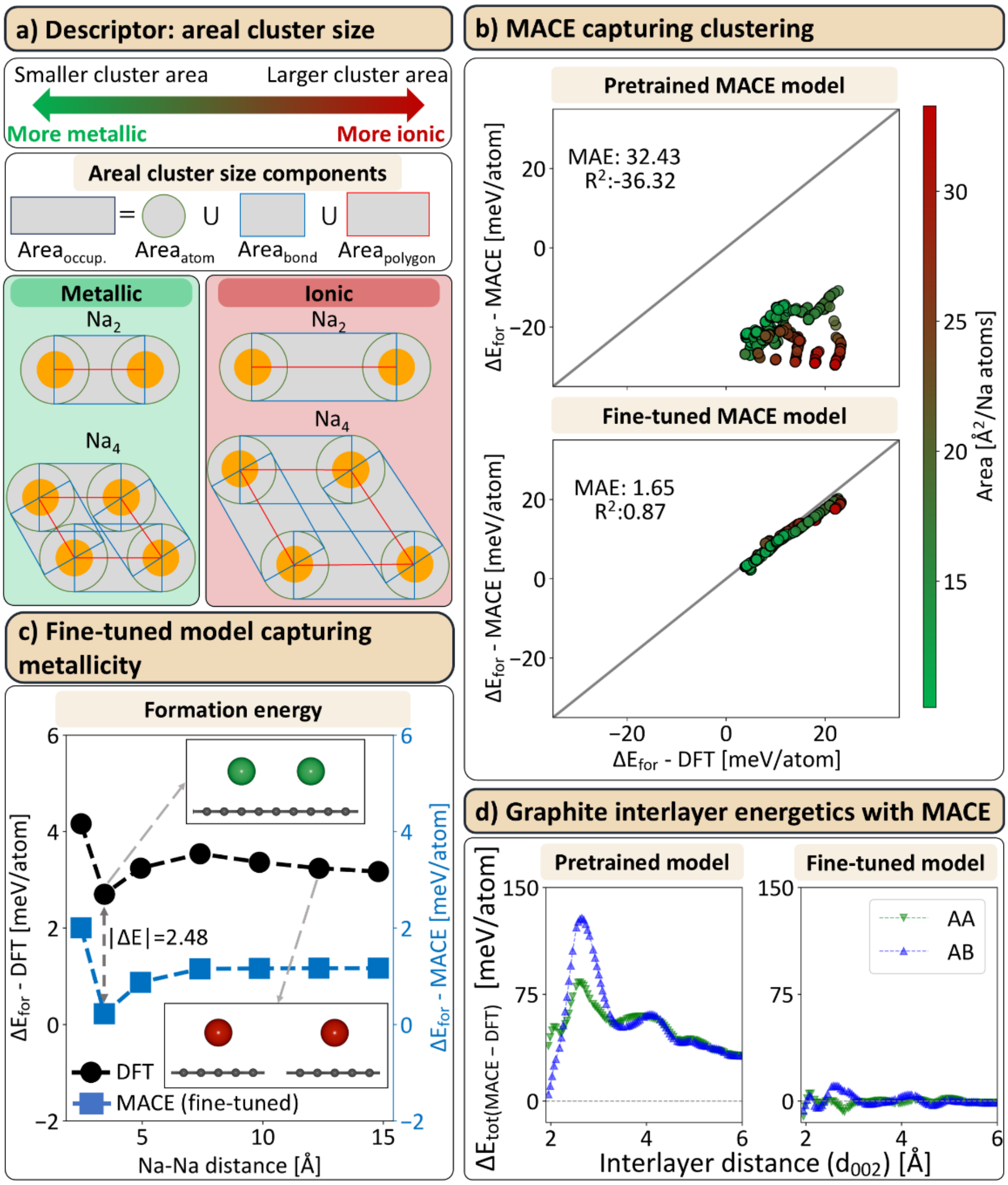

**Figure 2. Benchmarks for the proof of concept of the fine-tuned MACE model for Na clustering in graphene.** (a) Cluster area as a descriptor to analyze two-dimensional Na clusters; a smaller area per atom corresponds to Na clustering, while a larger area corresponds to dispersed sodium. (b) Comparison of the pretrained and fine-tuned MATPES models for the clustering tendency of different Na clusters on the graphene surface (only Na clusters are included on graphene); energetic values without dispersion. (c) Application of the fine-tuned model for the metallic interaction of two Na atoms on the graphene surface; energetic values are without dispersion. (d) Reference-state dependence of the difference in total energies between MACE and DFT as a function of the interlayer distance on graphite with AA and AB stacking.

Having established that the missing Na-clustering and graphite-reference motifs can be corrected by targeted fine-tuning, we next extend the model beyond this minimal *'proof-of-concept'* toward a broader Na/C configurational space. This is particularly important as hard carbon contains a broader range of environments, including graphitic domains, defects, amorphous regions, and pore-like regions. Motivated by this, we compiled a DFT database of these motifs (i.e., '*broader-dataset*'), consisting of Na clusters on graphene (combining planar and compressed Na clusters), Na clusters on graphene with a monovacancy, Na intercalation in graphite, pseudo-graphitic carbon, amorphous Na/C structures, and expanded references of graphite AA/AB and sodium BCC (Figs. 3a and 3b; see Methods in the SI). To compare the motif span in our dataset with that used to pretrain the MACE model, we extracted all structures containing only C, only Na, and Na/C-based systems from the MATPES dataset. Using the structures of our dataset and those extracted from MATPES, we calculated and applied the Ewald Sum Matrix descriptor[33] to capture structural information. With the calculated geometrical descriptors per structure, we reduced the dimensionality in a two-step process: first, using the TruncatedSVD algorithm in scikit-learn[34] to obtain a 50-dimensional array, and then using the UMAP algorithm[35] to reduce it to a 2-dimensional array. Using the reduced 2D array, we compared our dataset with MATPES (Fig. 3b), showing that the MATPES dataset has limited coverage of motifs such as Na arrangements on pristine graphene and graphene with monovacancies (MV). Furthermore, even in regions where MATPES captures intercalation, our dataset includes more configurations, indicating richer sampling of Na/C-based systems. Finally, on the points related to pseudographitic carbon, as we observed in Fig. 3b, the pretrained model does not include enough systems to represent the graphitic domain of carbon. Using this broader dataset (Fig. 3b), we fine-tuned the final large Na/C MACE model following the workflow illustrated in Fig. 3a, starting from the pretrained MATPES model. As illustrated in this workflow, a key point in for fine-tuning the model is filtering by atomic forces; as we can observe in Fig. S9, the amorphous dataset introduces high forces, and the distribution of structures depends on the cutoff for atomic forces (atomic forces obtained from DFT); with few structures with larger atomic forces per component ($|F_{i,\alpha\ atom}| > 20$ eV Å$^{-1}$ with $\alpha \in \{x, y, z\}$), their presence during the fine-tuning produces an increase in outliers that disproportionally bias the loss ; for this reason, the compiled broader dataset needs to be filtered. Fig. 3c shows the training-set performance for the model fine-tuned with structures filtered by $|F_{i,\alpha\ atom}| < 20$ eV Å$^{-1}$. Although the targeted foundation model was fine-tuned with structures with $|F_{i,\alpha\ atom}| < 20$ eV Å$^{-1}$, most of these structures have $|F_{i,\alpha\ atom}| < 5$ eV Å$^{-1}$, as illustrated in Fig. S10. Based on this distribution and the fact that the fine-tuned model will not be used for simulations with high forces, we select the range of $|F_{i,\alpha\ atom}| < 10$ eV Å$^{-1}$ as the intended application range for this fine-tuned foundation model. The model accurately reproduces the DFT formation energies, with an MAE of 10.67 meV atom$^{-1}$ and $R^2 \approx 1.00$, and also describes the force components with an MAE of 0.023 eV Å$^{-1}$ and $R^2 = 0.99$. We choose the model trained with a cutoff of 20 eV Å$^{-1}$, as it presents the lowest RMSEs for formation energies and atomic forces per component among the fine-tuned models (see Figs. S11 and S12). These results indicate that the final targeted fine-tuned model can consistently fit the diverse Na/C environments in the training set, ranging from compact Na clusters and defective graphene to intercalated graphite and amorphous carbon-based structures. Comparing both fine-tuned foundation models, the one obtained in the *'proof-of-concept'* (trained only using part of the Na clustering dataset and references of sodium BCC and graphite) with the *'broader-dataset'* (i.e., latest model) (trained with the broader compiled dataset), we can observe that the *'proof-of-concept'* fine-tuned model yields larger errors for predicted formation energies and atomic forces for the broader dataset (see Figs. S13 and S14), with MAE of 17.16 meV atom$^{-1}$ and $R^2$ of 0.99 for the formation energies in the training set, while for the atomic forces present a MAE of 0.065 eV Å$^{-1}$ and $R^2$ = 0.95 (accuracy metrics calculated for structures filtered by $|F_{i,\alpha\ atom}|$ < 10 eV Å$^{-1}$). As expected, like the pretrained MATPES model, it misses Na/C chemistries, such as amorphous structures and sodium adsorbed on graphene with monovacancies (see Fig. 3b).

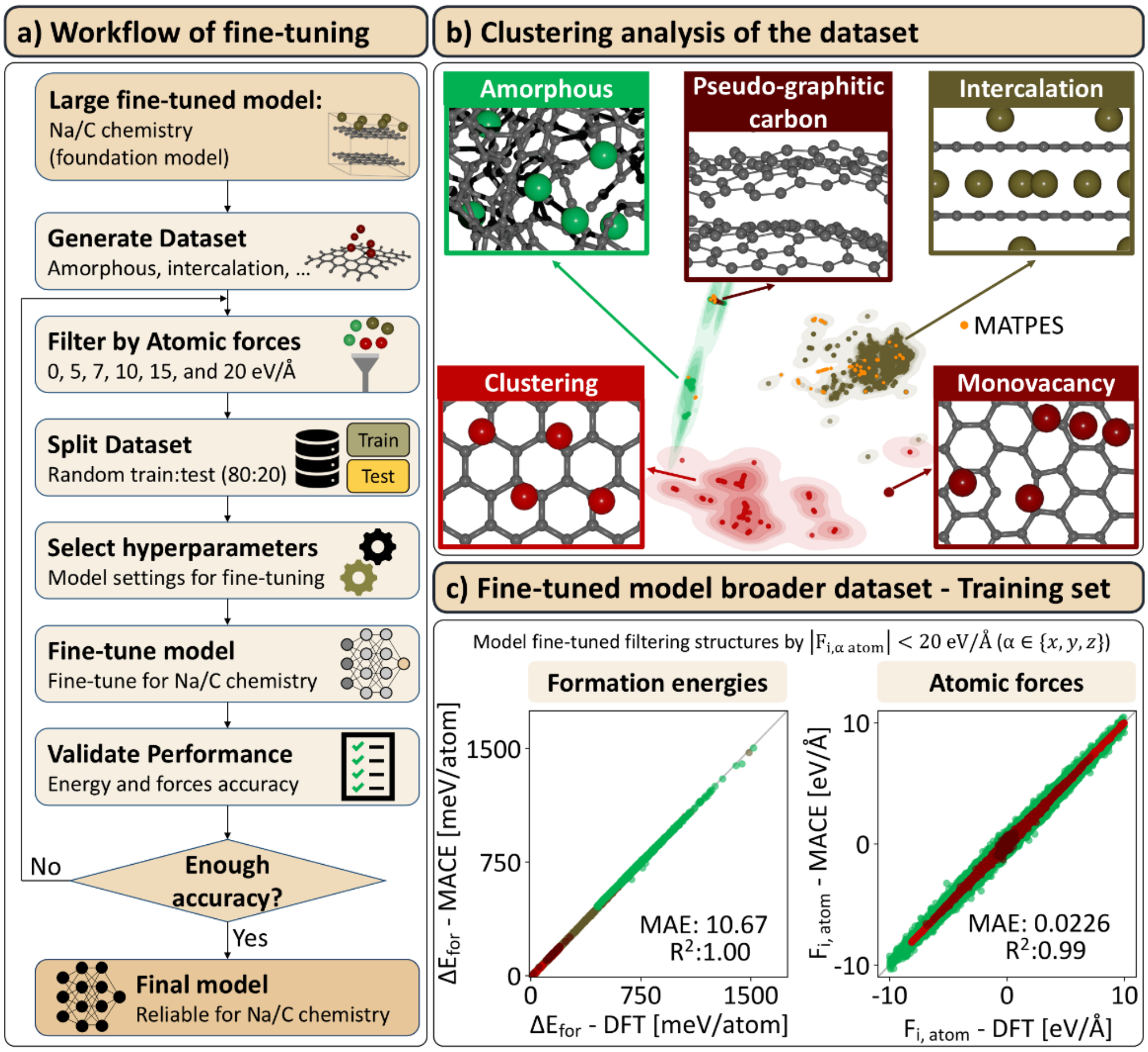

**Figure 3. Fine-tuned MACE capturing different C/Na energetic regions.** (a) Workflow to fine-tune the MATPES model with the broader dataset. (b) Clustering analysis comparison of the configurational region captured in the pretrained MATPES model with the configurational region captured in the dataset used in this work (dataset filtered with $|F_{i,\alpha\ atom}|$ < 20 eV/Å for fine-tuning). The contour is to show the density distribution per cluster, but does not represent the boundaries per cluster; the descriptor is the Ewald Sum Matrix reduced to 2D with the UMAP dimensionality reduction algorithm. (c) Parity plot for the training set: (left) formation energies and (right) atomic force per component for the dataset (without expanded graphite AA/AB and sodium) used in this work (for visualization and to calculate accuracy metrics, structures with $|F_{i,\alpha\ atom}|$ < 10 eV/Å are used); energetic values are without van der Waals dispersion.

It is important to note that the performance of the targeted fine-tuned foundation model with the *'broader-dataset'* in Fig. 3 is only for the training set (80% of the compiled dataset). Therefore, to verify that the model does not simply memorize the fine-tuning structures, we evaluated it on configurations not used during fitting, as summarized in Fig. 4a. Specifically, first, we tested the model within the same Na/C chemical domain using the test set obtained from the random train/test split of the compiled dataset. For this test set, we compared the results again after filtering structures by $|F_{i,\alpha\ atom}| < 10$ eV Å$^{-1}$ (atomic force per component); the model keeps nearly the same accuracy as for the training set: the formation-energy MAE is 10.70 meV atom$^{-1}$ with $R^2 \approx 1.00$, while the force-component MAE is 0.023 eV Å$^{-1}$ with $R^2 = 0.99$ (Fig. 4b). The similarity between the errors for the training (Fig. 3c) and test sets indicates that the model is not only fitting individual structures, but can interpolate reliably within the sampled Na/C configurational space. These similarities between the training and test sets are consistently observed when each specific subset is analyzed separately (see Figs. S15-S24). Fine-tuning the MATPES model with the *'broader-dataset'*also improves the prediction of the energetic stability of Na/C structures associated with the intercalation mechanism of Na storage (see Fig. 1b). For the intercalation subset from the compiled dataset, the formation-energy MAE is 9.99 meV atom$^{-1}$ with $R^2 \approx 1.00$ for the training set (Fig. S15), while the formation-energy MAE is 10.05 meV atom$^{-1}$ with $R^2 \approx 1.00$ for the test set (Fig. S20). Although this increase in performance was expected because the broader compiled dataset contains structures related to intercalation (Fig. 3b), the similar training and test errors (which are smaller than those of the pretrained MATPES models) further support the reliability of the targeted fine-tuned model with the broader dataset. For the local motifs related to the adsorption of Na arrangements, we can observe that in Fig. S17 and Fig. S22, the results show a poor $R^2$, suggesting that the fine-tuned model with the broader compiled dataset loses the description of the clustering of Na; however, first, there are some points to highlight: in these figures, compared to Fig. 2b, the dataset of clustering is larger, containing compressed Na clusters on pristine graphene (this subset was included to describe the shorter Na-Na bond distances; see more detailed information in the SI). Additionally, for the adsorption mechanism, we also include Na adsorbed on graphene monovacancies (a subset generated by molecular dynamics; see more detail in the SI), as shown in Figs. S18 and S23, where $R^2$ is higher, indicating a better fit for this motif. Then, to clarify why the lower $R^2$ is observed in Figs. S17 and S22, we used the dataset generated for the *'proof-of-concept'* (planar Na clusters with different Na-Na bond distances) and compared the fine-tuned model with the broader dataset and the pretrained MATPES model, as illustrated in Fig. S25; from the comparison of Na arrangements, we observed that the formation energies for the fine-tuning have the same trend as the values obtained from DFT: the more compressed clusters are more stable than the dispersed ones, but there is a constant shifting from the diagonal of the parity plot, which in part could be produced by the reference states (graphite AB and sodium BCC). Additionally, comparing our fine-tuned model with the existing pretrained models, we observed that the fine-tuned model presents higher accuracy for the formation energies and atomic forces (see Figs. S26-S33). In particular, for the pretrained MATPES model (see Fig. S26-S27) with both training and test sets, the energy-related MAE is around 25 meV atom$^{-1}$, almost two times that obtained with the fine-tuned model (~10.70 meV atom$^{-1}$). For the other pretrained model, the error is higher, whereas notably the fine-tuned model has the smallest error for the atomic forces, which is important for searching for the global minimum and the most thermodynamically favorable structures. Also, another important aspect of this fine-tuned model compared to the *'proof-of-concept'* is that it does not only capture a single chemical environment for Na/C; we introduced several different local motifs to balance the dataset and capture as broad a range as possible, meaning that this does not have high precision for a specific region (i.e., adsorption), but in general it can capture all the regions with good precision.

To further evaluate the transferability of the fine-tuned model, we applied the model to Na-filled carbon nanotubes (CNTs) (Fig. 4c,d), geometries which were excluded from the broader compiled dataset discussed in Fig. 3b. The dataset samples possible configurations of Na-filled CNTs with different diameters using the genetic algorithm implemented in the Atomic Simulation Environment (ASE).[36] For this, we used three CNTs: CNT(15,0), CNT(16,0), and CNT(17,0) (modeled by one unit cell), performing relaxation calculations with DFT to obtain the convex hull. The convex hull allows us to analyze the most thermodynamically stable phase for a Na/C morphology, in this case for Na-CNT. The genetic algorithm was used to allow us to iteratively generate new structures based on perturbation operations to increase the screening of unexplored structures to get as close as possible to the global minimum (see Fig. S34). Even for this Na-CNT dataset, the targeted fine-tuned model reproduces DFT formation energies with an MAE of 2.03 meV atom$^{-1}$ and $R^2$ = 0.97, and force components with an MAE of 0.054 eV Å$^{-1}$ and $R^2$ = -6.92. In the case of the negative atomic force component error, $R^2$ may reflect small variance relative to the reference forces from DFT; this is attributed to the fact that, for DFT, many of these points are close to zero (relaxed structures). Additionally, for this Na-CNT dataset, we compared the results of formation energies obtained via DFT with all the pretrained models, as shown in Figs. S35-S37, we can observe that the fine-tuned model presents the lowest MAE. Because of this, it can be concluded that the final fine-tuned model retains accuracy outside the training set and can transfer the learned Na-C/Na-Na interactions to structures outside of the training dataset. For the application of this fine-tuned model, one concern is related to how well the thermodynamic stability is described; in this sense, we use the formation energy as a descriptor, but specifically for the CNTs, we can observe that the formation energies are very small. To analyze how sensitive the predicted energies are to these small values, we took the structures predicted with the genetic algorithm by DFT. First, we used the reference state energies (i.e., each empty CNT and sodium BCC) obtained by DFT (see Fig. S38) and compared them with the reference state energies obtained by MACE (see Fig. S39). From this comparison, we can observe that the energies for the Na-filled CNTs predicted by the fine-tuned model are lower than those obtained by DFT; however, due to the reference state energies from MACE, the structures are less stable. Due to the differences between forces from DFT and the fine-tuned model, it is expected that there are differences in the most stable structures; we calculate the convex hull for those CNTs (using one unit cell) using the fine-tuned model (*'broader-dataset'*), aligned with the prediction of large forces (see Fig. 4d);

the ground states predicted by DFT and MACE are different (see Fig. S40). However, the formation energies of CNT(15,0) and CNT(16,0) are relatively close, with a difference below 0.20 meV/atom; for CNT(17,0), this difference is larger, with the fine-tuned model predicting more stable structures than DFT. These differences in energies are important because they will also affect the prediction of other properties, such as the voltage. For CNT(15,0) and CNT(16,0), we observe a voltage difference of around 1 mV (see Fig. S40), whereas for CNT(17,0) it is larger. By comparing the ground-state structural changes predicted by DFT and MACE, we analyze changes in bond lengths and angles (see Fig. S41). In general, the most affected features are the predicted carbon bond lengths and angles, which is partially expected because our compiled dataset includes no motifs related to empty CNTs or highly curved structures.

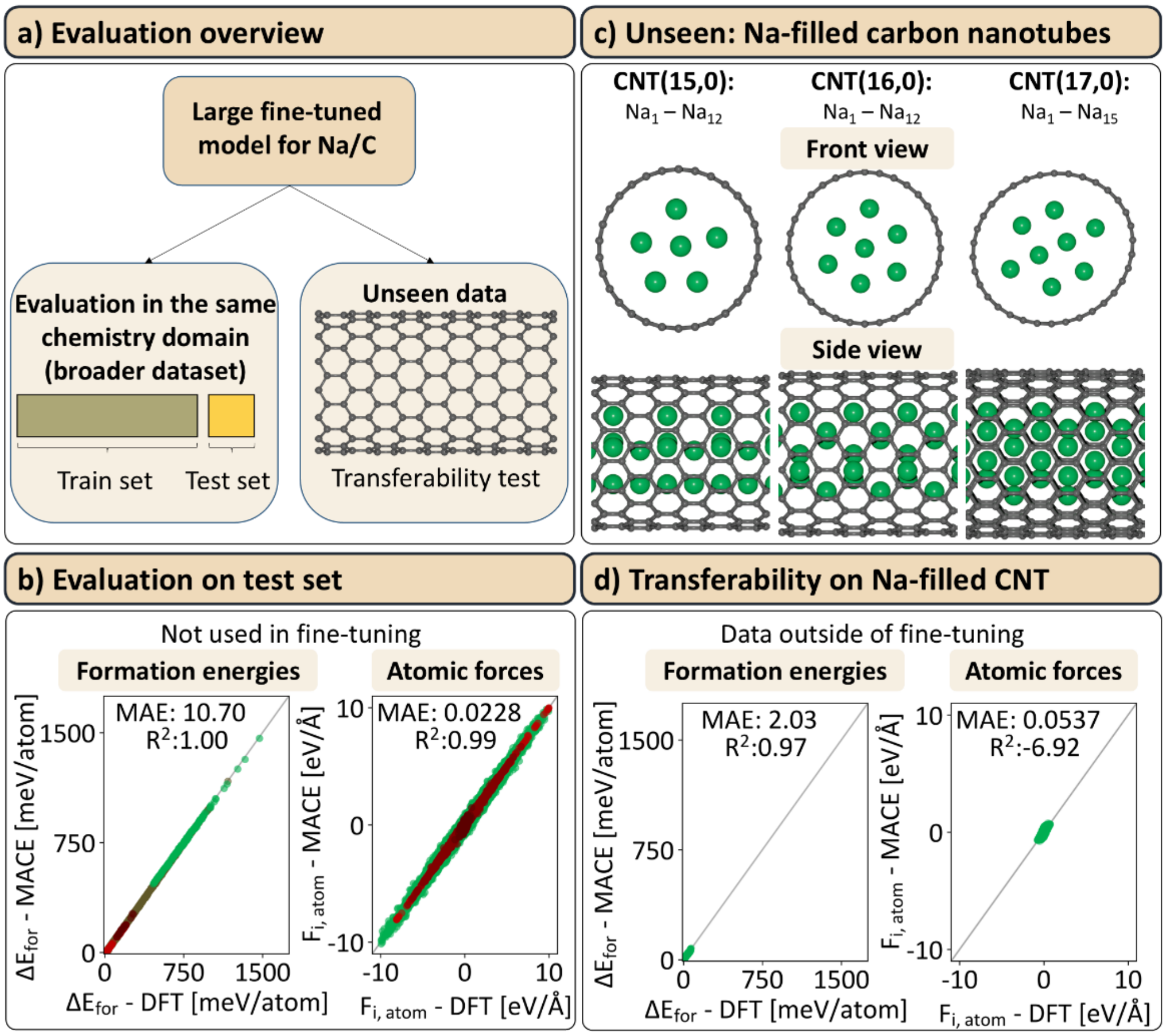

**Figure 4. Transferability of the fine-tuned model with the broader compiled dataset.** (a) Schematic description of the dataset used to evaluate the performance of the targeted fine-tuned model. (b) Parity plot for the test set (without expanded graphite AA/AB and sodium): (left) formation energies and (right) atomic forces per component (for visualization and to calculate accuracy metrics, structures with $|F_{i,\alpha\ atom}| < 10$ eV/Å are used); energetic values are without van der Waals dispersion. (c) Front and side views of the most stable structures for Na filling different CNTs. (d) Parity plot for CNTs filled with sodium: CNT(15, 0) from $Na_1$ to $Na_{12}$, CNT(16, 0) from $Na_1$ to $Na_{12}$, and CNT(17, 0) from $Na_1$ to $Na_{15}$: (left) formation energies and (right) atomic force per component (dataset not filtered by maximum atomic force); for this dataset, dispersion is included via the D3-BJ correction.

Once we verified that the targeted fine-tuned model captures Na-C/Na-Na chemistry and, more importantly, recovers the stability of different Na/C environments, we delved into the Na storage mechanism. From the analysis of the training and test sets for the broader compiled dataset, we observed that because the dataset contains Na/C chemical environments related to adsorption and intercalation, the fine-tuned model captures the energetic stability of these local motifs, but one part is missing for a broad analysis of the Na storage mechanism in hard carbon: pore filling. One key aspect of the Na storage mechanism, especially for the plateau voltage, is the filling of closed pores [10,27–29] which is the most critical for commercial application. Our representation for closed pores is idealized as a cylindrical geometry consisting of a pristine periodic CNT with four unit cells; for this case, we used CNT(15,0), CNT(17,0), CNT(19,0), CNT(21,0), and CNT(23,0) with radii of 0.59, 0.67, 0.75, 0.83, and 0.91 nm, respectively. Compared with the DFT calculations illustrated in Fig. 4d, these simulations for the closed pores involve larger systems (> 240 C atoms), but this is not just to evaluate how the fine-tuned model performs with a larger system; it is also important because of a technical limitation of DFT: a shorter-cell-size DFT simulation may constrain Na arrangements and omit the prediction of the structures with the lowest energy in the convex hull. An energetic analysis was performed to calculate the convex hull via the genetic algorithm, as shown in Fig. 5a. Using each respective empty closed pore (CNT) and sodium BCC as references, we observe that in the convex hull for our closed-pore models, only one structure is stable per closed pore; however, there are several metastable configurations with formation energies below zero but lying above the convex hull, meaning that they are thermodynamically favorable with respect to the empty closed pore and sodium BCC references but are not the most stable phase. These configurations could potentially be stabilized by thermal effects or kinetically trapped under operating conditions. Additionally, these metastable configurations could act as intermediates on the path to the most stable configuration with the highest Na capacity. For closed pores with larger radii, such as r equal to 0.83 and 0.91 nm, the absence of thermodynamically favorable low-loading configurations for the larger pores suggests that filling may require a higher Na chemical potential, or the formation of sufficiently large Na arrangements before stabilization of the most stable phase occurs. As a final analysis, we compared how each closed pore contributes to the voltage profile; as expected, with larger radii, the total capacity increases (Fig. 5b); however, each closed pore contributes at a different plateau voltage, which implies that in a real hard carbon system, where instead of a homogeneous shape of closed pores, a distribution of different pores is present, each pore contributes differently. Moreover, this implies heterogeneous filling: closed pores with a higher driving force (smaller radius) fill first, and some large pores may not reach maximum capacity under operating conditions.

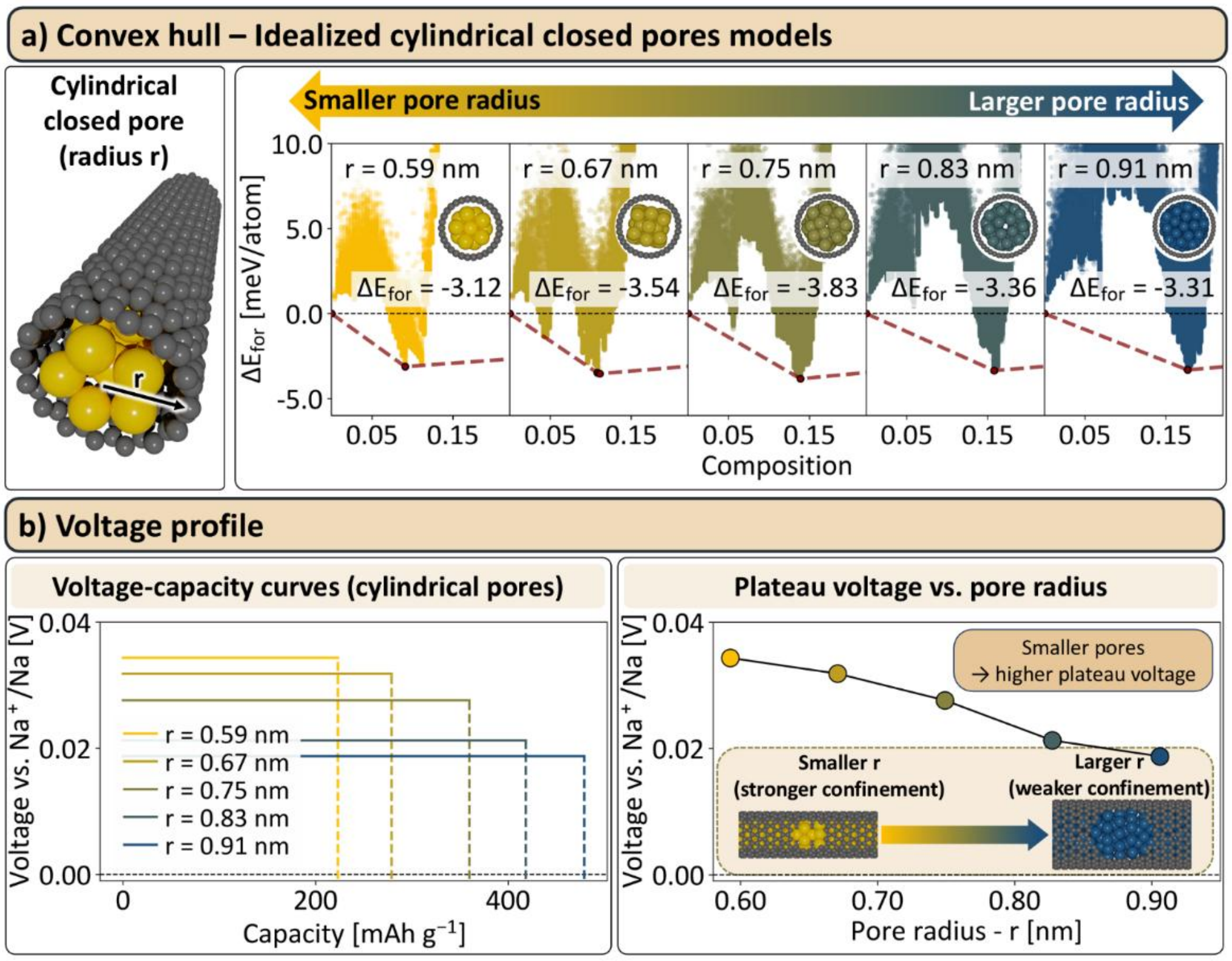

**Figure 5. Application of the fine-tuned model: influence of the radius of pores on Na storage in closed pores.** (a) Convex hull for idealized cylindrical closed pores predicted with the genetic algorithm. (b) Voltage profile: (left) average voltage vs. gravimetric capacity, (right) plateau voltage as a function of the radius. For all the energetic analyses, the references were each respective empty closed pore (depending on the radius) and sodium BCC using the fine-tuned model trained with the broader compiled dataset; for all the energetic analyses, the D3-BJ dispersion corrections were included.

## CONCLUSIONS

Using sodium storage in hard carbon as a challenging example of complex chemical environments, we systematically analyzed local motifs in hard carbon and described several local bonding interactions. We showed that the pretrained models can fail to capture the energetic competition between Na-C interaction and Na-Na cohesion, which is essential to describe the Na storage mechanism in carbon-based systems. We show that targeted fine-tuning can recover the energetic description of metallic clusters adsorbed on graphene, and furthermore, with fine-tuning on a broader dataset, we show that the model presents enough transferability to capture Na/C chemistries excluded from the dataset used to fine-tune it. These findings show that global accuracy metrics tell us about how well the model fits; in the case of pretrained foundation models, it is also important to conduct a deeper exploration and evaluation, which helps to understand how well the atomic interactions are captured. This analysis and fine-tuning of the foundation model with the broader compiled dataset illustrate that this model can recover the energetic description of Na-C and Na-Na bonding compared with the pretrained foundation models. We also explore the ability of the fine-tuned model to link atomistic findings with experimental design parameters of hard carbon. Using a pristine cylindrical model to represent the pore and study pore filling, we perform a thermodynamic analysis using the convex hull and voltage profile for electrode insertion; we observe that the thermodynamic driving force is increased for the smallest radius, which also implies higher voltages, while the larger pores can accommodate a larger number of sodium atoms but at lower voltages. This correlation between pore radius and voltage suggests that each pore geometry contributes differently to the voltage profile, and instead of a unique plateau voltage, we present a voltage contribution as a function of the pore-size distribution. More importantly, we show that foundation models can construct heuristic rules for materials design from a bottom-up approach, but this requires not only mimicking the energetic signature of local chemistries but also capturing the physics of competing bonding regimes.

## METHODS

All DFT calculations were performed under the framework implemented in the Vienna ab initio Simulation Package (VASP)[37–39] with the projector augmented wave (PAW) method[40] and the Perdew-Burke-Ernzerhof (PBE) functional.[41] Different computational setups (VASP parameters) were used to generate the structures that constitute the compiled dataset. Calculations such as relaxation and Ab Initio Molecular Dynamics (AIMD) were performed, and, in the case of the structures of Na intercalated in graphite from our previous work,[6] a different computational setup was used; however, once all the structures were generated, Self-Consistent Field (SCF) calculations were performed using a computational setup identical to that of the MATPES dataset but without spin-polarization.[30] For dispersion, the D3-BJ damping correction was used, but none of the calculations in the compiled dataset used to fine-tune the MATPES model included this correction.[42,43] Atomistic foundation models based on the MACE architecture were used for this work,[15,23] extracting all reported-to-date pretrained models. For fine-tuning the pretrained MATPES model, among the different fine-tuning methods for MACE, we chose the replay multi-head method,[44] which was used to conserve part of the original learning of the model. Our approach consists of fine-tuning the MATPES with two different weights (two-step process), with 100 epochs in total, where the first 40 epochs have a higher weight for reducing forces and stress (1:10:10 energy:force:stress, respectively), and in the last 60 epochs the weight of the energy was increased (10:10:10 energy:force:stress, respectively). A more detailed description of the selection of the pretrained model to fine-tune, methodology, and generation of the dataset can be found in the Supplementary Information and references therein.[6,13,30,31,33,35–41,44–52]

## ACKNOWLEDGEMENTS

The work in China was supported by the Startup Fund of Qingyuan Innovation Laboratory (Grant No. 00524009), the Fujian Province High-Level Innovative and Entrepreneurial Talents Program (Grant No. 30225003), the National Key Research and Development Program of China (Grant No. 2024YFE0101600), and Natural Science Foundation of Fujian Province (Grant No. 2026I0007). The work in Poland was supported by the National Centre for Research and Development under the project WPC3/2022/50/KEYTECH/2024. Institutional and infrastructural support for the ENSEMBLE3 Centre of Excellence was provided through the 2.1 International Research Agendas programme of the Foundation for Polish Science co-financed by the European Union under the European Funds for Smart Economy 2021-2027 (FENG.02.01-IP.05-0044/24), project (MAB/2020/14), which is carried out within the International Research Agendas Programme (IRAP) of the Foundation for Polish Science co-financed by the European Union under the European Regional Development Fund and the Teaming Horizon 2020 programme (GA. No. 857543) of the European Commission and the project of the Minister of Science and Higher Education "Support for the activities of Centers of Excellence established in Poland under the Horizon 2020 program" under contract No. MEiN/2023/DIR/3797. We acknowledge the Polish high-performance computing infrastructure, PLGrid, for granting this project access to computational resources under project number PLG/2026/019890, as well as to the LUMI supercomputer, owned by the EuroHPC Joint Undertaking and hosted by CSC (Finland) and the LUMI consortium, through project number PLL/2025/09/018974.

## CONFLICT OF INTEREST

The authors declare that they have no known competing financial interests or personal relationships that could have appeared to influence the work reported in this paper.

## DATA AVAILABILITY STATEMENT

All data necessary to reproduce the results are available within the article and its Supplementary Information. The underlying calculation files and computational codes are available from the corresponding author upon reasonable request.

## AUTHOR CONTRIBUTIONS

**A.F.U.:** Data curation, Formal analysis, Visualization, Writing – original draft, Writing – review & editing, Methodology, Investigation, Software. **I.R.:** Writing – review & editing, Investigation, Software. **Y.T.:** Writing – review & editing, Funding acquisition. **O.I.M.:** Writing – original draft, Writing – review & editing, Resources, Project administration, Methodology, Investigation, Funding acquisition, Conceptualization.